\documentclass[longbibliography,amsmath,amssym,amsthm,superscriptaddress,twocolumn]{revtex4-2}
\usepackage[colorlinks=true,urlcolor=blue,citecolor=blue,linkcolor=blue]{hyperref}
\usepackage{graphicx}
\usepackage{color}
\usepackage{amsfonts}
\usepackage{hyperref}
\usepackage{newfloat}
\usepackage{bm}
\usepackage[utf8]{inputenc}

\newcommand{\msf}{\mathsf}

\begin{document}
\title{Estimating entropy production in a stochastic system with odd-parity variables}
\author{Dong-Kyum \surname{Kim}}
\thanks{Equal contribution.}
\affiliation{Department of Physics, Korea Advanced Institute of Science and Technology, Daejeon 34141, Korea}
\author{Sangyun \surname{Lee}}
\thanks{Equal contribution.}
\affiliation{School of Physics, Korea Institute for Advanced Study, Seoul 02455, Korea}
\author{Hawoong \surname{Jeong}}
\email{hjeong@kaist.edu}
\affiliation{Department of Physics, Korea Advanced Institute of Science and Technology, Daejeon 34141, Korea}
\affiliation{Center for Complex Systems, Korea Advanced Institute of Science and Technology, Daejeon 34141, Korea}

\begin{abstract}
Entropy production (EP) is a central measure in nonequilibrium thermodynamics, as it can quantify the irreversibility of a process as well as its energy dissipation in special cases. Using the time-reversal asymmetry in a system's path probability distribution, many methods have been developed to estimate EP from only trajectory data. However, estimating the EP of a system with odd-parity variables, which prevails in nonequilibrium systems, has not been covered. In this study, we develop a machine learning method for estimating the EP in a stochastic system with odd-parity variables through multiple neural networks. We demonstrate our method with two systems, an underdamped bead-spring model and a one-particle odd-parity Markov jump process.
\end{abstract}

\maketitle
\section{Introduction}

Entropy production (EP) plays an important role in the field of stochastic thermodynamics because of its diverse relations to the dissipation, free energy, and fundamental bounds on the fluctuation of an observable. Through the fluctuation relation~\cite{seifert2005entropy}, it has been found that EP has a symmetry that enables us to measure the free energy difference from nonequilibrium experimental data~\cite{Collin2005verify}. Also, trade-off relations between EP and the relative error of physical quantities have been discovered recently~\cite{barato2015thermodynamic, Proesmans_2017, Hasegawa2019, Koyuk2020}. Such relations can provide a tighter version of the thermodynamic second law and a universal bound for fluctuations of diverse currents.  

Following from the importance, various estimation methods for EP have been developed, such as the plug-in method~\cite{Roldan2010Estimating} and compression algorithm~\cite{Avinery2019Universal}. Also, the development of thermodynamic uncertainty relations (TURs) open the possibility of measuring the EP in overdamped Langevin processes~\cite{Gingrich2016Dissipation}. To estimate the exact EP with TURs, we have to measure the relative error of EP, which presents us with circular reasoning. To resolve this point, a few optimization strategies have been established~\cite{Li2019quantifying, Vu2020Entropy,otsubo2020estimating,dechant2021improving}. However, measuring EP remains challenging because of the curse of dimensionality. 

Recently, machine learning based algorithms for EP estimation have been developed to efficiently mitigate such issues by using a neural network~\cite{kim2020neep,otsubo2020time,bae2021attaining}. These methods employ neural networks as an EP estimator~\cite{kim2020neep}, where the estimator is trained for optimizing the variational representation of the EP~\cite{donsker1983varadhan}. Owing to the great power of neural networks, these approaches work well even in high-dimensional systems. Moreover, unlike the TURs methods, the neural estimator can provide stochastic EP directly from trajectory data without any correction~\cite{otsubo2020time}.

Despite such progress, the developed methods have only considered systems with absent odd-parity variables, of which the sign changes under time-reversal operation~\cite{Richard2012Nonequilibrium, lee2013fluctuation, lee2021universal}. For instance, underdamped Langevin dynamics includes an odd-parity variable called momentum~\cite{Risken}; in many cases, active matter or other biological systems are often modeled with underdamped dynamics~\cite{shankar2018hidden,Dabelow2019Irreversibility}. To estimate the EP of such systems, existing methodologies require additional information because the definition of EP in these systems differs from that in systems without odd-parity variables. Indeed, TURs for underdamped Langevin systems require detailed information about the systems~\cite{vu2019tur, Lee2019Thermodynamic, lee2021universal}, and therefore we cannot exactly measure EP from only trajectory data using TURs. Due to these difficulties, to our knowledge, methods for estimating EP in a stochastic system with odd-parity state variables have yet to be developed.

In this article, we propose a machine learning based method for estimating EP by using trajectory data obtained from a stochastic system with odd-parity variables. The proposed method can be applied to Langevin dynamics and Markov jump processes. For a system with continuous state variables, we demonstrate our estimation method with an underdamped bead-spring model in Sec.~\ref{sec:ubs}, and for the Markov jump case, we apply our approach in Sec.~\ref{sec:omj} to a one-particle odd-parity Markov jump process in which the state variables are discrete and the time interval between jumps is continuous.

\section{Underdamped bead-spring model}
\label{sec:ubs}

In the underdamped bead spring model, $N$ beads are coupled to the next nearest beads. Let $\bm{q}=(x_1, \dots, x_N, v_1, \dots, v_N)$ be a state vector of this model, and its time reversal counterpart is denoted as $\widetilde{\bm{q}}=(x_1, \dots, x_N, -v_1, \dots, -v_N)$ where $x_i$ and $v_i$ are the position and velocity of the $i$-th bead, respectively. Then, the dynamics of the state $\bm{q}$ with time $\tau$ is governed by the following underdamped Langevin equation:
\begin{align}
    \dot{\bm{q}}(\tau) = - \msf{M} \cdot \bm{q}(\tau) + \sqrt{2\msf{D}} \cdot \bm{\xi}(\tau),
\label{eq:ubs}
\end{align}
with 
\begin{align}
    \msf{M}=\begin{pmatrix}
    \msf{0} & -\msf{I} \\
    \msf{A} & \msf{\Gamma}
    \end{pmatrix}, \quad \msf{D}=\begin{pmatrix}
    \msf{0} & \msf{0} \\
    \msf{0} & \msf{T}
    \end{pmatrix},
\end{align}
where $[\msf{A}]_{ij}= (2\delta_{i,j} - \delta_{i, j+1} - \delta_{i+1, j})k/m $, $[\msf{\Gamma}]_{ij}=\gamma \delta_{i,j}$, and $[\msf{T}]_{ij}= \delta_{i,j} \gamma k_{\rm B} T_i / m $. Otherwise, $\bm{\xi}$, is a Gaussian noise vector with zero mean and variance given by $\langle \xi_i(\tau)\xi_j(\tau')\rangle =\delta_{ij}\delta(\tau-\tau')$, and $\langle \dots \rangle$ denotes the ensemble average. $\msf{0}$ and $\msf{I}$ are the $N \times N$ null and identity matrices, $\gamma$ is the friction coefficient, and $k$ is a spring constant. We set $k=k_{\rm B}=\gamma=1$~\footnote{Here, we note that $[\msf{\Gamma}]_{ij}$ is not $\delta_{i,j}\gamma/m$ but $\delta_{i,j}\gamma$. This means that the governing equation is not reduced to an overdamped Langevin equation when mass $m$ approaches zero because the velocity autocorrelation is proportional to $e^{-\tau\gamma}$ rather than $e^{-\tau\gamma/m}$.}.

For data preparation, we sample an ensemble of trajectories from Eq.~\eqref{eq:ubs}. $\bm{q}_1$ is sampled from steady-state distribution (SSD) $p(\bm{q})$, and we use the second-order integrator to sample a trajectory $(\bm{q}_1, \bm{q}_2, \dots, \bm{q}_L)$, where $\bm{q}_t \equiv \bm{q}(t\Delta\tau)$ with fixed sampling interval $\Delta \tau$~\cite{vanden2006second}. In this case, the stochastic EP per step is defined as 
\begin{align}
\Delta S_{\rm tot}(\bm{q}_t, \bm{q}_{t+1}) = \ln{\frac{p(\bm{q}_{t+1} | \bm{q}_t)}{p(\widetilde{\bm{q}}_t | \widetilde{\bm{q}}_{t+1})}} + \ln{\frac{p(\bm{q}_{t})}{p(\bm{q}_{t+1})}},
\label{eq:ubs_ep}
\end{align}
where the first (second) term on the right-hand side is the environment EP (system entropy change)~\cite{seifert2012stochastic}.

In order to estimate Eq.~\eqref{eq:ubs_ep}, we employ the objective function proposed in Ref.~\cite{kim2020neep}:
\begin{align}
J_{\theta_1} = \langle \Delta S_{\theta_1}(\bm{q}_t, \bm{q}_{t+1}) - e^{-\Delta S_{\theta_1}(\bm{q}_t, \bm{q}_{t+1})} + 1 \rangle_t,
\label{eq:ubs_obj_1}
\end{align}
where
\begin{align*}
\Delta S_{\theta_1}(\bm{q}_t, \bm{q}_{t+1}) = h_{\theta_1}(\bm{q}_t, \bm{q}_{t+1}) - h_{\theta_1}(\widetilde{\bm{q}}_{t+1}, \widetilde{\bm{q}}_t),
\end{align*}
and $\langle \dots \rangle_t$ denotes the expectation over $t$ and the ensemble. Here, we use a multi-layer perceptron (MLP) for $h_{\theta_1}$, where $\theta_1$ represents the trainable neural network parameters. Using the stochastic gradient ascent method, we train the neural network $h_{\theta_1}$ to find the optimal parameter $\theta_1^{\star}$ which gives the maximum value of Eq.~\eqref{eq:ubs_obj_1}. The optimal parameter $\theta_1^{\star}$ gives
\begin{align} 
    \Delta S_{\theta_1^{\star}}(\bm{q}_t, \bm{q}_{t+1}) = \ln{\frac{p(\bm{q}_t, \bm{q}_{t+1})}{p(\widetilde{\bm{q}}_{t+1}, \widetilde{\bm{q}}_t)}},
\end{align}
and this can be split into
\begin{align*}
    \Delta S_{\theta_1^{\star}}(\bm{q}_t, \bm{q}_{t+1}) =\ln{\frac{p(\bm{q}_{t+1}|\bm{q}_t)}{p(\widetilde{\bm{q}}_{t} | \widetilde{\bm{q}}_{t+1})}} + \ln{\frac{p(\bm{q}_t)}{p(\bm{q}_{t+1})}}+ \ln{\frac{p(\bm{q}_{t+1})}{p(\widetilde{\bm{q}}_{t+1})}} .
\end{align*}
Therefore, the maximum value of $J_{\theta_1}$ is
\begin{align}
    J_{\theta_1^{\star}} &= \left< \ln{\frac{p(\bm{q}_{t+1}|\bm{q}_t)}{p(\widetilde{\bm{q}}_{t} | \widetilde{\bm{q}}_{t+1})}} + \ln{\frac{p(\bm{q}_t)}{p(\bm{q}_{t+1})}} + \ln{\frac{p(\bm{q}_{t+1})}{p(\widetilde{\bm{q}}_{t+1})}}\right>_t  \\
    &= \langle \Delta S_{\rm tot}(\bm{q}_t, \bm{q}_{t+1})\rangle_t + \left<\ln{\frac{p(\bm{q})}{p(\widetilde{\bm{q}})}}\right>.
    \label{eq:optimal_j1}
\end{align}
The last term of Eq.~\eqref{eq:optimal_j1} is the Kullback--Leibler divergence between the SSD $p(\bm{q})$ and its mirror state distribution $p(\widetilde{\bm{q}})$. Equation~\eqref{eq:optimal_j1} can be written as 
\begin{align}
    J_{\theta_1^{\star}} = \dot{\sigma}\Delta \tau + \left<\Delta S_{\rm as}(\bm{q})\right>,
\label{eq:ubs_obj_1_sol}
\end{align}
where $\dot{\sigma}$ is the EP rate and $\Delta S_{\rm as}(\bm{q})$ is the asymmetry in SSD-given state $\bm{q}$:
\begin{align}
    \Delta S_{\rm as}(\bm{q}) = \ln{\frac{p(\bm{q})}{p(\widetilde{\bm{q}})}}.
    \label{eq:s_as}
\end{align}
To measure the EP rate $\dot{\sigma}$, we need to estimate $\Delta S_{\rm as}(\bm{q})$. To do so, we define another objective function:
\begin{align}
J_{\theta_0} = \langle \Delta S_{\theta_0}(\bm{q}_t) - e^{-\Delta S_{\theta_0}(\bm{q}_t)} + 1 \rangle_t,
\label{eq:ubs_obj_0}
\end{align}
where $\Delta S_{\theta_0}(\bm{q}_t) = h_{\theta_0}(\bm{q}_t) - h_{\theta_0}(\widetilde{\bm{q}}_t)$ and $h_{\theta_0}$ is another MLP. Using the same method as before, we maximize Eq.~\eqref{eq:ubs_obj_0}.
The optimal parameter $\theta_0^{\star}$ gives 
\begin{align}
\Delta S_{\theta_0^{\star}}(\bm{q}_t) &= \ln{\frac{p(\bm{q}_t)}{p(\widetilde{\bm{q}_t})}}, \\ 
J_{\theta_0^{\star}} &= \langle \Delta S_{\rm as}(\bm{q}) \rangle.
\label{eq:ubs_obj_0_sol}
\end{align}
Using the two trained neural estimators, the estimations of the stochastic EP $\Delta S_{\rm tot}$ and EP rate $\dot{\sigma}$ are, respectively, 
\begin{align}
    \Delta\hat{S}_{\rm tot}(\bm{q}_t, \bm{q}_{t+1}) = \Delta S_{\theta_1}(\bm{q}_t, \bm{q}_{t+1}) - \Delta S_{\theta_0}(\bm{q}_{t+1}),
\label{eq:ubs_stochastic_ep}
\end{align}
and 
\begin{align}
    \hat{\dot{\sigma}} = \frac{J_{\theta_1} - J_{\theta_0}}{\Delta\tau}.
\label{eq:ubs_ep_rate}
\end{align}

\begin{figure*}[!t]
\centering
\includegraphics[width=\textwidth]{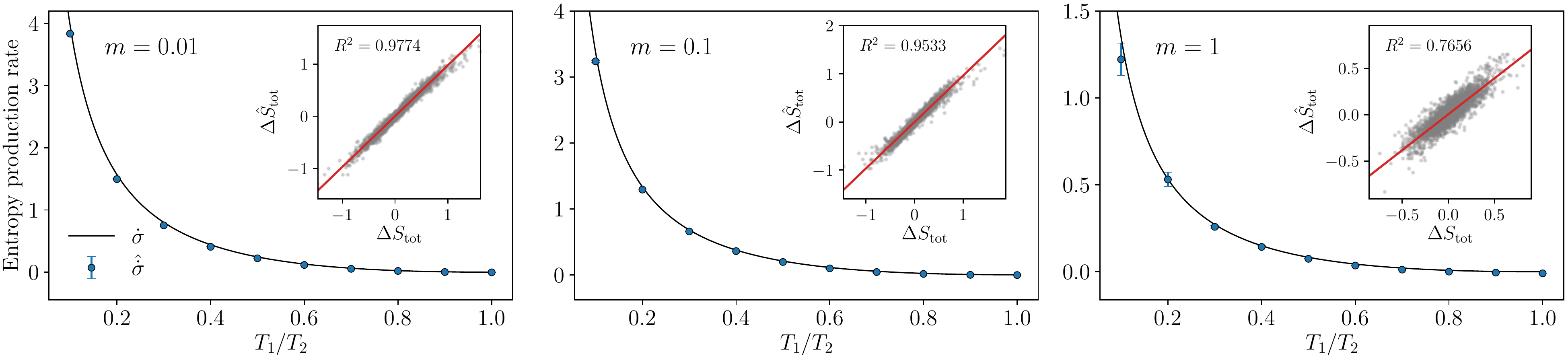}
\vskip -0.1in
\caption{Entropy production rate as a function of $T_1/T_2$ in the underdamped bead-spring model for $m=0.01$ (left), $0.1$ (middle), and $1$ (right). The solid lines (blue dots) represent the analytical EP rate~$\dot \sigma$ (estimated EP rate~$\hat{\dot \sigma}$). The insets are scatter plots of the analytical~$\Delta S_{\rm tot}$ and the estimation~$\Delta \hat{S}_{\rm tot}$ at $T_1/T_2=0.1$, where the red lines denote linear regression. Error bars represent the standard deviation of $\hat{\dot{\sigma}}$ from five independently trained estimators.}
\label{fig:ubs}
\vskip -0.1in
\end{figure*}

To verify our approach, we sample the training, validation, and test set data with sampling interval $\Delta \tau=0.01$, trajectory length $L=4000$, ensemble size (number of trajectories) $M=10^4$, and number of beads $N=2$. We fix the temperature of the second bead to $T_2=10$, and the datasets are sampled from $T_1 \in \{1, 2, \dots, 10\}$, each with three different values of mass $m\in \{0.01, 0.1, 1.0\}$. 

We train two MLPs, $h_{\theta_0}$ and $h_{\theta_1}$, to maximize Eq.~\eqref{eq:ubs_obj_1} and Eq.~\eqref{eq:ubs_obj_0}, respectively, by using the Adam optimizer~\cite{kingma2014adam} with learning rate $10^{-5}$ and batch size 4096. Each MLP has a single output unit and two hidden layers of 256 units with rectified linear units (ReLU) nonlinearity. The parameters of these MLPs are initialized with the orthogonal initialization scheme~\cite{saxe2014ortho}, which is good for stable and efficient convergence. We use this setup for all training processes unless otherwise noted. As a data preprocessing step, we normalize each position and velocity in state $\bf{q}$ before feeding to the MLPs. The total number of training iterations is $10^5$, and we evaluate $J_{\theta_0}$ and $J_{\theta_1}$ values from the validation set every 400 training iterations. The best trained parameters $\{\theta_0^\star, \theta_1^\star\}$ are obtained by monitoring the $J_{\theta_0}$ and $J_{\theta_1}$ values during the training processes. Throughout this work, all the results in the figures are evaluated from the test set, where the error bars represent the standard deviation of the estimations from five independently trained estimators.

In Fig.~\ref{fig:ubs}, we plot the training results for estimating EP in the underdamped bead-spring model. For $m=0.01$ and $m=0.1$ (left and middle plot), the estimated EP rates $\hat{\dot{\sigma}}$ (blue dots) are well matched with the analytic EP rates $\dot{\sigma}$ (solid lines) with small variance. The insets in Fig.~\ref{fig:ubs} are scatter plots between the estimated stochastic EP $\Delta\hat{S}_{\rm tot}$ and analytic stochastic EP $\Delta S_{\rm tot}$ with a fitted linear regression line (red line), showing that the stochastic EP is also well estimated with high $R^2$ values. However, the training result for $m=1$ (right plot) shows high variance in the EP rate estimation at $T_1/T_2=0.1$, and the stochastic EP is inaccurately estimated with $R^2=0.7656$.

\begin{figure}[!t]
\centering
\includegraphics[width=\columnwidth]{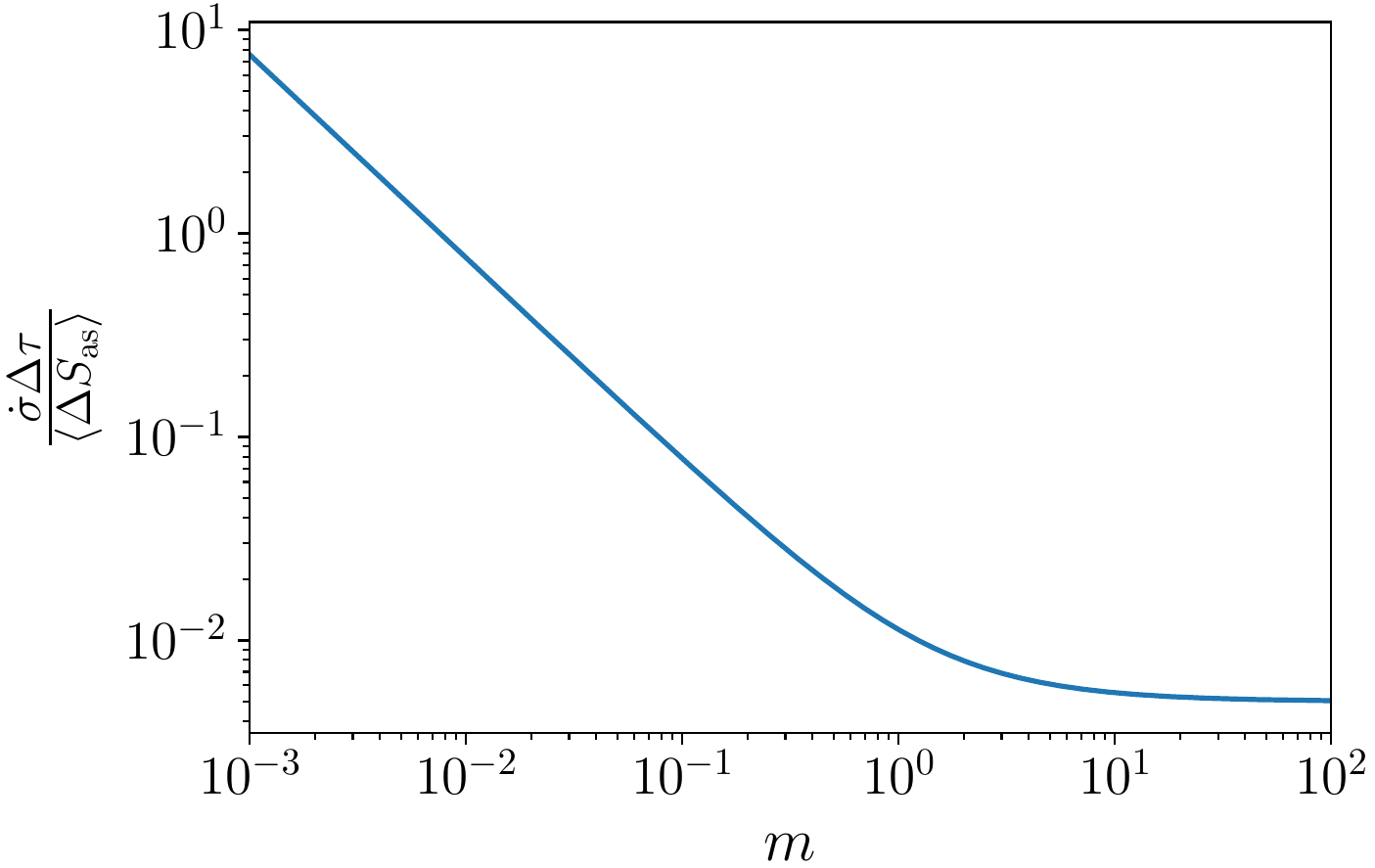}
\vskip -0.1in
\caption{Entropy production per step over $\left<\Delta S_{\rm as}\right>$ as a function of mass $m$ at $T_1/T_2=0.1$.}
\label{fig:ubs_asym}
\vskip -0.1in
\end{figure}

In order to identify these inaccuracies, in Fig.~\ref{fig:ubs_asym} we compare the analytic value of $\langle\Delta S_{\rm as}\rangle$ and EP per step $\dot{\sigma}\Delta \tau$ while changing the mass $m$ at $T_1/T_2=0.1$. As can be seen in the figure, the total EP per step is much larger than $\langle\Delta S_{\rm as}\rangle$ in the small $m$ regime, but with increasing $m$, the total EP per step becomes smaller than $\langle\Delta S_{\rm as}\rangle$. At $m=1$, the ratio $\dot{\sigma}\Delta \tau / \langle\Delta S_{\rm as}\rangle$ is almost 0.01; such a small ratio leads to a very slight difference between Eq.~\eqref{eq:ubs_obj_1_sol} and Eq.~\eqref{eq:ubs_obj_0_sol}, which may contribute to the difficulty in EP estimation from Eq.~\eqref{eq:ubs_ep_rate} that optimizes $J_{\theta_1}$ and $J_{\theta_0}$ separately.

\section{One-particle odd-parity Markov jump process}
\label{sec:omj}

Next, we consider a one-particle Markov jump process on a ring with $N$ sites~\cite{lee2013fluctuation}. In this model, the particle state is defined as $\bm q=(n,v)$, and its time reversal state is denoted as $\widetilde{\bm q}=(n,-v)$ where $n\in\{1,\dots,N\}$ is the position variable and $v\in\{-1,+1\}$ is velocity variable. The master equation is given as 
\begin{align}
    \partial_t p(\bm q, t) =  \sum_{\bm q'\neq \bm q } [\omega_{\bm q, \bm q'} p(\bm q',t) - \omega_{\bm q', \bm q} p(\bm q,t) ],
\label{eq:Markov_jump}\end{align}
where $\omega_{\bm q',\bm q}$ denotes the transition rate for the jump from $\bm q$ to $\bm q'$, and the transition rates are given as $\omega_{(n\pm v,\pm v),(n,\pm v)} = f_{\pm v}$ and $\omega_{(n,\mp v),(n,\pm v)} = r_{\pm v}$.
For a jump of waiting time $\tau$ ($\bm q \rightarrow \bm q'$, $\tau$), the entropy production~\cite{lee2013fluctuation, martinez2019inferring} is written as 
\begin{align}
    \Delta{S}_{\rm tot}(\bm q,\bm q',\tau) =& \ln{\frac{\omega_{\bm q',\bm q} e^{-\tau/\tau_{\bm q}} }{\omega_{\tilde{\bm q},\tilde{\bm  q}'} e^{-\tau/\tau_{\widetilde{\bm q} }}}}   + \ln{ \frac{p(\bm q)}{p(\bm q')} },
\label{eq:omj_ep1_true}
\end{align}
where $p(\bm q)\equiv p(\bm q, \infty)$ is the steady state of Eq.~\eqref{eq:Markov_jump}. The rightmost term in Eq.~\eqref{eq:omj_ep1_true} is the system EP of the steady state.
The first term on the right-hand side denotes the EP due to the jump and the wait.
The numerator~$\omega_{\bm q',\bm q} \psi_{\bm q}(\tau)$ is the transition rate $\bm q \rightarrow \bm q'$ after waiting for time $\tau$ in the state~$\bm q$. The denominator~$\omega_{\tilde{\bm q},\tilde{\bm  q}'} e^{-\tau/\tau_{\widetilde{\bm q} }}$ is the corresponding reversed transition rate with $\tau$ waiting time in the state~$\widetilde{\bm q}$ after a jump~$\widetilde{\bm q}' \rightarrow \widetilde{\bm q}$. The mean waiting time of the state $\bm q$ is $\tau_{\bm q}=1/\sum_{\bm q'\neq \bm q}\omega_{\bm q',\bm q}$. 

For a Markov jump process, we represent a trajectory~$\Gamma$ as a sequence of jumps: 
\begin{align}
    \Gamma = \{({\bm q}_1 \rightarrow  {\bm q}_2, \tau_1),\dots, ({\bm q}_i\rightarrow {\bm q}_{i+1}, \tau_i), \dots\},
\label{eq:Markov_traj}
\end{align}
where $q_i$ denotes the $i$-th state. To train neural networks with the sequence of jumps, we rearrange Eq.~\eqref{eq:Markov_traj} into
\begin{align}
\begin{split}
    \Delta{S}_{\rm tot}(\bm q,\bm q',\tau) 
    =& \ln{ \frac{R(\widetilde{\bm{q} }' )}{R(\bm q') } } + \ln{ \frac{R(\widetilde{\bm{q} } )}{R(\bm q) } }\\ 
    & +\ln{\frac{\rho_{\bm q',\bm q}R(\bm q)}{{\rho}_{\widetilde{\bm q}, \widetilde{\bm q}'}R(\widetilde{\bm q}')}}
    + \ln{\frac{\psi_{\bm q}(\tau) R({\bm q}) }{\psi_{\widetilde{\bm q} }(\tau)R(\widetilde{\bm q})}}\\
    &+\ln{\frac{\tau_{\bm q}\tau_{\widetilde{\bm q}'}}{\tau_{\widetilde{\bm q}},\tau_{\bm q'}}}
\end{split}\label{eq:omj_ep_true}\\
    \equiv& \Delta{S}_{\rm alt}(\bm q, \bm{q}', \tau) +\ln{\frac{\tau_{\bm q}\tau_{\widetilde{\bm q}'}}{\tau_{\widetilde{\bm q}},\tau_{\bm q'}}},
\end{align}
where $\rho_{\bm q',\bm q} \equiv \tau_{\bm q} \omega_{\bm q',\bm q}$ denotes the jump probability from $\bm q$ to $\bm q'$ and $R_{\bm q} \equiv \tau^{-1}_{\bm q} p(\bm q)/\sum_{\bm q'} \tau^{-1}_{\bm q'}p(\bm q')$ is the stationary probability distribution of jump probabilities $\rho_{\bm q',\bm q}$~\cite{martinez2019inferring}. 
The waiting time distribution (WTD) in the state~$\bm q$ is denoted by $ \psi_{\bm q}(\tau) \equiv \frac{e^{-\tau/\tau_{\bm q}}}{\tau_{\bm q}}$. 
The last term of Eq.~\eqref{eq:omj_ep_true}, which consists of mean waiting times, can be directly measured from without any estimator hence we focus on estimating $\Delta{S}_{\rm alt}(\bm q, \bm{q}', \tau)$. Because the average of the last term is zero, an averaged total entropy production and an averaged alternative entropy production are the same~$\langle \Delta S_{\rm tot}\rangle = \langle \Delta S_{\rm alt} \rangle$.
To estimate the stochastic EP [Eq.~\eqref{eq:omj_ep_true}], we require three kinds of estimators: $\Delta S_{\theta_0}$, $\Delta S_{\theta_1}$, and $\Delta S_{\theta_{2}}$.
The first and second terms on the right-hand side of Eq.~\eqref{eq:omj_ep_true} can be estimated with $\Delta S_{\theta_{0}}(\bm q)$ and $\Delta S_{\theta_{0}}(\bm q')$, respectively.
The third term can be estimated with $\Delta S_{\theta_{1}}(\bm q,\bm q')$.
We use the same objective functions as the previous section, Eq.~\eqref{eq:ubs_obj_0} and Eq.~\eqref{eq:ubs_obj_1}, for training $\Delta S_{\theta_0}$ and $\Delta S_{\theta_1}$, but we develop a different kind of estimator with a new objective function for estimating the fourth term on the right-hand side of Eq.~\eqref{eq:omj_ep_true}:
\begin{align}
J_{\theta_{2}} = \langle \Delta S_{\theta_{2}}(\bm q, \tau) - e^{-\Delta S_{\theta_{2}}(\bm q, \tau)} + 1 \rangle_{\tau},
\label{eq:omj_obj_wtd}
\end{align}
where 
\begin{align}
\Delta S_{\theta_{2}}(\bm q, \tau) = h_{\theta_{2}}(\bm q, \tau) - h_{\theta_{2}}(\widetilde{\bm q}, \tau).
\end{align}
The optimal parameter $\theta_{2}^{\star}$ gives
\begin{align}
\Delta S_{\theta_{2}^{\star}}(\bm q, \tau) &= \ln{\frac{\psi_{\bm q}(\tau)}{\psi_{\widetilde{\bm q}}(\tau)}} + \ln{\frac{R_{\bm q}}{R_{\widetilde{\bm q}}}},
\label{eq:third_ent}\\
J_{\theta_{2}^{\star}} &= \langle \Delta S_{\theta_{2}}(\bm q, \tau) \rangle_{\tau} + \langle \Delta S_{\rm as}(\bm q) \rangle.
\end{align}
The first term of Eq.~\eqref{eq:third_ent} is the contribution from the waiting time $\tau$ to the EP. We denote this contribution as 
\begin{align}
    \Delta{S}_{\rm WTD}(\bm q, \tau) \equiv \ln{\frac{\psi_{\bm q}(\tau)}{\psi_{\widetilde{\bm q}}(\tau)}}.
\label{eq:s_wtd}
\end{align}
Then, using these three trained neural estimators, we can estimate the EP $\Delta S_{\rm alt}$ and EP rate $\dot{\sigma}$ with the following equations:
\begin{align}
\begin{split}
    \Delta{S}_{\rm alt}(\bm q, \bm{q}', \tau)& = \Delta S_{\theta_1}(\bm q, \bm{q}') + \Delta S_{\theta_{2}}(\bm q, \tau) \\
    &- \Delta S_{\theta_0}(\bm q) - \Delta S_{\theta_0}(\bm{q}'),
\label{eq:omj_ep}
\end{split}
\end{align}
and 
\begin{align}
    \hat{\dot{\sigma}} = \frac{J_{\theta_1} + J_{\theta_{2}}- 2J_{\theta_0}}{\tau_{\rm avg}},
\label{eq:omj_ep_rate}
\end{align}
where $\tau_{\rm avg}$ denotes the average of the waiting time $\tau$ from the trajectory data. To handle the discrete state variables, we replace the input layers of $h_{\theta_0}$, $h_{\theta_1}$, and $h_{\theta_{2}}$ with an embedding layer that converts discrete values into continuous space vectors. The dimension of the embedding layer is 128; other neural network configurations and training setup are the same as in the previous section.

\begin{figure}[!t]
\centering
\includegraphics[width=\columnwidth]{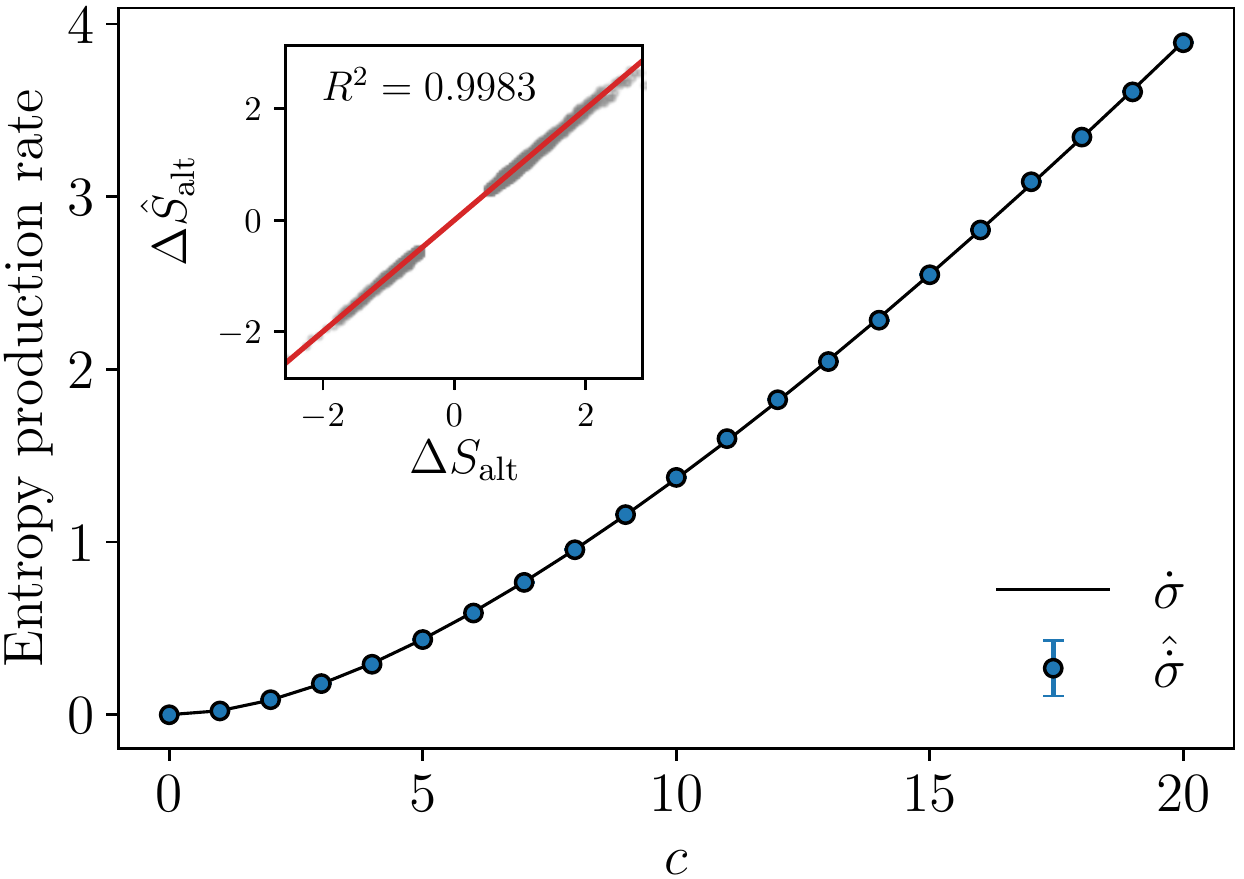}
\vskip -0.1in
\caption{Entropy production rate as a function of $c$ in the one-particle odd-parity Markov jump process. The inset is a scatter plot of $\Delta S_{\rm tot}$ and $\Delta \hat{S}_{\rm tot}$ at $c=10$ where the red line denotes the linear regression. The standard deviation (error bar) of the EP rate estimation $\hat{\dot{\sigma}}$ is much smaller than the symbol size.}
\label{fig:omj}
\vskip -0.1in
\end{figure}

To validate this method, we sample an ensemble of trajectories with total number of jumps $L=10^4$, ensemble size $M=50$, and number of sites $N=10$. The Gillespie algorithm~\cite{gillespie1977exact} is used for sampling a sequence of $L-1$ jumps,
\begin{align*}
    \Gamma = \{({\bm q}_1 \rightarrow  {\bm q}_2, \tau_1), \dots, ({\bm q}_{L-1}\rightarrow {\bm q}_L, \tau_{L-1})\}.
\end{align*}
We set the transition rates as $f_v=r_{-v}=1$, $f_{-v}=1+0.1c$, and $r_{v}=1+0.2c$, where $c$ is a positive real number. Each of the training, validation, and test trajectory datasets are sampled with $c$ in the 0--20 range. When $c=0$, the steady state of this system is in equilibrium. Non-zero~$c$ induces thermodynamic currents and increases EP in the steady state.

We train three neural networks  $h_{\theta_0}$, $h_{\theta_1}$, $h_{\theta_{2}}$ to maximize $J_{\theta_0}$, $J_{\theta_1}$, and $J_{\theta_{2}}$, respectively. The total training iterations are $10^4$, and $J_{\theta_0}$, $J_{\theta_1}$, and $J_{\theta_{2}}$ are evaluated every 500 iterations from the validation dataset. As shown in Fig.~\ref{fig:omj}, the EP rate estimation $\hat{\dot{\sigma}}$ [Eq.~\eqref{eq:omj_ep_rate}] agrees with the analytic EP rate $\dot{\sigma}$. A scatter plot between $\Delta S_{\rm tot}$ and $\Delta \hat{S}_{\rm tot}$ [Eq.~\eqref{eq:omj_ep}] at $c=10$ with a fitted linear regression line (red line) is shown in the Fig.~\ref{fig:omj} inset. Overall $R^2$ values of the linear regression between $\Delta S_{\rm alt}$ and $\Delta \hat{S}_{\rm alt}$ are above 0.99.

\section{Discussion}
\label{sec:discussion}
Unlike systems with only even-parity variables, in which EP can be estimated with a single neural estimator~\cite{kim2020neep}, we require additional estimators to treat the mirror state asymmetry $\Delta S_{\rm as}$ [Eq.~\eqref{eq:s_as}] and waiting-time contribution $\Delta S_{\rm WTD}$ [Eq.~\eqref{eq:s_wtd}] in the estimation of EP in systems with odd-parity variables.

The first example in this work demonstrated that EP in the underdamped bead-spring model can be estimated by two neural estimators $\Delta S_{\theta_0}$ and $\Delta S_{\theta_1}$ which are trained to maximize $J_{\theta_0}$ and $J_{\theta_1}$, respectively. However, this method has a limitation when $J_{\theta_0}$ and $J_{\theta_1}$ values are very close to each other; a small difference between these two values may cause a high variance in the EP estimations. This problem comes the training of two independent objective functions $J_{\theta_0}$ and $J_{\theta_1}$. As shown in Eq.~\eqref{eq:ubs_obj_0_sol} and Eq.~\eqref{eq:ubs_obj_1_sol}, $J_{\theta_1}$ includes contribution from $\langle \Delta S_{\rm as}\rangle$ [Eq.~\eqref{eq:s_as}], so if the $\langle S_{\rm as}\rangle$  contribution in $J_{\theta_1}$ is dominant, then estimating the EP from the difference between $J_{\theta_0}$ and $J_{\theta_1}$ might be hard. Also, $\hat{\dot{\sigma}}$ [Eq.~\eqref{eq:ubs_ep_rate}] is not a lower bound on EP rate $\dot{\sigma}$. While $J_{\theta_0}$ and $J_{\theta_1}$ are the lower bounds on Eq.~\eqref{eq:ubs_obj_0_sol} and Eq.~\eqref{eq:ubs_obj_1_sol}, respectively, this does not guarantee that $\hat{\dot{\sigma}}=(J_{\theta_1} - J_{\theta_1})/\Delta \tau$ is a lower bound on the EP rate. Therefore, designing a single objective function that can bound the EP will be an important future work.

The second example demonstrated the estimation of the EP in an odd-parity Markov jump process with three neural estimators $\Delta S_{\theta_0}$, $\Delta S_{\theta_1}$, and $\Delta S_{\theta_{2}}$ trained to maximize $J_{\theta_0}$, $J_{\theta_1}$, and $J_{\theta_{2}}$, respectively.
We stress that our method is the first tool to estimate EP in a Markov jump process with odd-parity variables.
As shown in Eq.~\eqref{eq:s_wtd}, the WTD does not contribute to the EP when the system has no odd-parity variables. However, if the accessible states are coarse-grained variables, then the WTD plays an important role despite the absence of odd-parity variables. In a related work, Martínez~{\it et al.}~\cite{martinez2019inferring} developed a method using WTD statistics to infer the EP from trajectories of coarse-grained states whose dynamics are described as semi-Markov processes; this method can infer EP even in the absence of observable currents. Also, a recent work by Skinner and Dunkel~\cite{skinner2021wtd} uses the fluctuation in WTD to infer the EP rate in the hidden Markov processes of two coarse-grained states. As a future work, developing an estimator for EP when odd-parity variables are coarse-grained will be an intriguing task, because the inferred EP from the trajectory of coarse-grained states can be larger than the actual EP when odd-parity variables are coarse-grained~\cite{kawaguchi2013hidden,esposito2015hidden}.

In summary, we have developed a machine learning based method that can estimate the stochastic EP and EP rate in a fluctuating system with odd-parity variables via neural networks. Our method was demonstrated with both an underdamped Langevin process and Markov jump process with odd-parity variables. It is expected to provide a practical tool for studying nonequilibrium systems in general experimental situations.

The results of all runs and the code implemented in \texttt{PyTorch}~\cite{pytorch2019} are available online~\footnote{\url{https://github.com/kdkyum/odd_neep}}.

\begin{acknowledgments} 
This study was supported by the Basic Science Research Program through the National Research Foundation of Korea (NRF) (KR) [NRF-2017R1A2B3006930 (D.-K.K., H.J.)], and a KIAS Individual Grant No. PG081801 (S. L.) from the Korea Institute for Advanced Study. 
\end{acknowledgments}


\begin{thebibliography}{37}%
\makeatletter
\providecommand \@ifxundefined [1]{%
 \@ifx{#1\undefined}
}%
\providecommand \@ifnum [1]{%
 \ifnum #1\expandafter \@firstoftwo
 \else \expandafter \@secondoftwo
 \fi
}%
\providecommand \@ifx [1]{%
 \ifx #1\expandafter \@firstoftwo
 \else \expandafter \@secondoftwo
 \fi
}%
\providecommand \natexlab [1]{#1}%
\providecommand \enquote  [1]{``#1''}%
\providecommand \bibnamefont  [1]{#1}%
\providecommand \bibfnamefont [1]{#1}%
\providecommand \citenamefont [1]{#1}%
\providecommand \href@noop [0]{\@secondoftwo}%
\providecommand \href [0]{\begingroup \@sanitize@url \@href}%
\providecommand \@href[1]{\@@startlink{#1}\@@href}%
\providecommand \@@href[1]{\endgroup#1\@@endlink}%
\providecommand \@sanitize@url [0]{\catcode `\\12\catcode `\$12\catcode
  `\&12\catcode `\#12\catcode `\^12\catcode `\_12\catcode `\%12\relax}%
\providecommand \@@startlink[1]{}%
\providecommand \@@endlink[0]{}%
\providecommand \url  [0]{\begingroup\@sanitize@url \@url }%
\providecommand \@url [1]{\endgroup\@href {#1}{\urlprefix }}%
\providecommand \urlprefix  [0]{URL }%
\providecommand \Eprint [0]{\href }%
\providecommand \doibase [0]{https://doi.org/}%
\providecommand \selectlanguage [0]{\@gobble}%
\providecommand \bibinfo  [0]{\@secondoftwo}%
\providecommand \bibfield  [0]{\@secondoftwo}%
\providecommand \translation [1]{[#1]}%
\providecommand \BibitemOpen [0]{}%
\providecommand \bibitemStop [0]{}%
\providecommand \bibitemNoStop [0]{.\EOS\space}%
\providecommand \EOS [0]{\spacefactor3000\relax}%
\providecommand \BibitemShut  [1]{\csname bibitem#1\endcsname}%
\let\auto@bib@innerbib\@empty
\bibitem [{\citenamefont {Seifert}(2005)}]{seifert2005entropy}%
  \BibitemOpen
  \bibfield  {author} {\bibinfo {author} {\bibfnamefont {U.}~\bibnamefont
  {Seifert}},\ }\bibfield  {title} {Entropy production along a stochastic
  trajectory and an integral fluctuation theorem,\ }\href
  {https://doi.org/10.1103/PhysRevLett.95.040602} {\bibfield  {journal}
  {\bibinfo  {journal} {Phys. Rev. Lett.}\ }\textbf {\bibinfo {volume} {95}},\
  \bibinfo {pages} {040602} (\bibinfo {year} {2005})}\BibitemShut {NoStop}%
\bibitem [{\citenamefont {Collin}\ \emph {et~al.}(2005)\citenamefont {Collin},
  \citenamefont {Ritort}, \citenamefont {Jarzynski}, \citenamefont {Smith},
  \citenamefont {Tinoco},\ and\ \citenamefont {Bustamante}}]{Collin2005verify}%
  \BibitemOpen
  \bibfield  {author} {\bibinfo {author} {\bibfnamefont {D.}~\bibnamefont
  {Collin}}, \bibinfo {author} {\bibfnamefont {F.}~\bibnamefont {Ritort}},
  \bibinfo {author} {\bibfnamefont {C.}~\bibnamefont {Jarzynski}}, \bibinfo
  {author} {\bibfnamefont {S.~B.}\ \bibnamefont {Smith}}, \bibinfo {author}
  {\bibfnamefont {I.}~\bibnamefont {Tinoco}},\ and\ \bibinfo {author}
  {\bibfnamefont {C.}~\bibnamefont {Bustamante}},\ }\bibfield  {title}
  {Verification of the Crooks fluctuation theorem and recovery of RNA folding
  free energies,\ }\href {https://doi.org/10.1038/nature04061} {\bibfield
  {journal} {\bibinfo  {journal} {Nature}\ }\textbf {\bibinfo {volume} {437}},\
  \bibinfo {pages} {231} (\bibinfo {year} {2005})}\BibitemShut {NoStop}%
\bibitem [{\citenamefont {Barato}\ and\ \citenamefont
  {Seifert}(2015)}]{barato2015thermodynamic}%
  \BibitemOpen
  \bibfield  {author} {\bibinfo {author} {\bibfnamefont {A.~C.}\ \bibnamefont
  {Barato}}\ and\ \bibinfo {author} {\bibfnamefont {U.}~\bibnamefont
  {Seifert}},\ }\bibfield  {title} {Thermodynamic uncertainty relation for
  biomolecular processes,\ }\href
  {https://doi.org/10.1103/PhysRevLett.114.158101} {\bibfield  {journal}
  {\bibinfo  {journal} {Phys. Rev. Lett.}\ }\textbf {\bibinfo {volume} {114}},\
  \bibinfo {pages} {158101} (\bibinfo {year} {2015})}\BibitemShut {NoStop}%
\bibitem [{\citenamefont {Proesmans}\ and\ \citenamefont {den
  Broeck}(2017)}]{Proesmans_2017}%
  \BibitemOpen
  \bibfield  {author} {\bibinfo {author} {\bibfnamefont {K.}~\bibnamefont
  {Proesmans}}\ and\ \bibinfo {author} {\bibfnamefont {C.~V.}\ \bibnamefont
  {den Broeck}},\ }\bibfield  {title} {Discrete-time thermodynamic uncertainty
  relation,\ }\href {https://doi.org/10.1209/0295-5075/119/20001} {\bibfield
  {journal} {\bibinfo  {journal} {{EPL} (Europhysics Letters)}\ }\textbf
  {\bibinfo {volume} {119}},\ \bibinfo {pages} {20001} (\bibinfo {year}
  {2017})}\BibitemShut {NoStop}%
\bibitem [{\citenamefont {Hasegawa}\ and\ \citenamefont
  {Van~Vu}(2019)}]{Hasegawa2019}%
  \BibitemOpen
  \bibfield  {author} {\bibinfo {author} {\bibfnamefont {Y.}~\bibnamefont
  {Hasegawa}}\ and\ \bibinfo {author} {\bibfnamefont {T.}~\bibnamefont
  {Van~Vu}},\ }\bibfield  {title} {Fluctuation Theorem Uncertainty Relation,\
  }\href {https://doi.org/10.1103/PhysRevLett.123.110602} {\bibfield  {journal}
  {\bibinfo  {journal} {Phys. Rev. Lett.}\ }\textbf {\bibinfo {volume} {123}},\
  \bibinfo {pages} {110602} (\bibinfo {year} {2019})}\BibitemShut {NoStop}%
\bibitem [{\citenamefont {Koyuk}\ and\ \citenamefont
  {Seifert}(2020)}]{Koyuk2020}%
  \BibitemOpen
  \bibfield  {author} {\bibinfo {author} {\bibfnamefont {T.}~\bibnamefont
  {Koyuk}}\ and\ \bibinfo {author} {\bibfnamefont {U.}~\bibnamefont
  {Seifert}},\ }\bibfield  {title} {Thermodynamic Uncertainty Relation for
  Time-Dependent Driving,\ }\href
  {https://doi.org/10.1103/PhysRevLett.125.260604} {\bibfield  {journal}
  {\bibinfo  {journal} {Phys. Rev. Lett.}\ }\textbf {\bibinfo {volume} {125}},\
  \bibinfo {pages} {260604} (\bibinfo {year} {2020})}\BibitemShut {NoStop}%
\bibitem [{\citenamefont {Rold\'an}\ and\ \citenamefont
  {Parrondo}(2010)}]{Roldan2010Estimating}%
  \BibitemOpen
  \bibfield  {author} {\bibinfo {author} {\bibfnamefont {E.}~\bibnamefont
  {Rold\'an}}\ and\ \bibinfo {author} {\bibfnamefont {J.~M.~R.}\ \bibnamefont
  {Parrondo}},\ }\bibfield  {title} {Estimating Dissipation from Single
  Stationary Trajectories,\ }\href
  {https://doi.org/10.1103/PhysRevLett.105.150607} {\bibfield  {journal}
  {\bibinfo  {journal} {Phys. Rev. Lett.}\ }\textbf {\bibinfo {volume} {105}},\
  \bibinfo {pages} {150607} (\bibinfo {year} {2010})}\BibitemShut {NoStop}%
\bibitem [{\citenamefont {Avinery}\ \emph {et~al.}(2019)\citenamefont
  {Avinery}, \citenamefont {Kornreich},\ and\ \citenamefont
  {Beck}}]{Avinery2019Universal}%
  \BibitemOpen
  \bibfield  {author} {\bibinfo {author} {\bibfnamefont {R.}~\bibnamefont
  {Avinery}}, \bibinfo {author} {\bibfnamefont {M.}~\bibnamefont {Kornreich}},\
  and\ \bibinfo {author} {\bibfnamefont {R.}~\bibnamefont {Beck}},\ }\bibfield
  {title} {Universal and Accessible Entropy Estimation Using a Compression
  Algorithm,\ }\href {https://doi.org/10.1103/PhysRevLett.123.178102}
  {\bibfield  {journal} {\bibinfo  {journal} {Phys. Rev. Lett.}\ }\textbf
  {\bibinfo {volume} {123}},\ \bibinfo {pages} {178102} (\bibinfo {year}
  {2019})}\BibitemShut {NoStop}%
\bibitem [{\citenamefont {Gingrich}\ \emph {et~al.}(2016)\citenamefont
  {Gingrich}, \citenamefont {Horowitz}, \citenamefont {Perunov},\ and\
  \citenamefont {England}}]{Gingrich2016Dissipation}%
  \BibitemOpen
  \bibfield  {author} {\bibinfo {author} {\bibfnamefont {T.~R.}\ \bibnamefont
  {Gingrich}}, \bibinfo {author} {\bibfnamefont {J.~M.}\ \bibnamefont
  {Horowitz}}, \bibinfo {author} {\bibfnamefont {N.}~\bibnamefont {Perunov}},\
  and\ \bibinfo {author} {\bibfnamefont {J.~L.}\ \bibnamefont {England}},\
  }\bibfield  {title} {Dissipation Bounds All Steady-State Current
  Fluctuations,\ }\href {https://doi.org/10.1103/PhysRevLett.116.120601}
  {\bibfield  {journal} {\bibinfo  {journal} {Phys. Rev. Lett.}\ }\textbf
  {\bibinfo {volume} {116}},\ \bibinfo {pages} {120601} (\bibinfo {year}
  {2016})}\BibitemShut {NoStop}%
\bibitem [{\citenamefont {Li}\ \emph {et~al.}(2019)\citenamefont {Li},
  \citenamefont {Horowitz}, \citenamefont {Gingrich},\ and\ \citenamefont
  {Fakhri}}]{Li2019quantifying}%
  \BibitemOpen
  \bibfield  {author} {\bibinfo {author} {\bibfnamefont {J.}~\bibnamefont
  {Li}}, \bibinfo {author} {\bibfnamefont {J.~M.}\ \bibnamefont {Horowitz}},
  \bibinfo {author} {\bibfnamefont {T.~R.}\ \bibnamefont {Gingrich}},\ and\
  \bibinfo {author} {\bibfnamefont {N.}~\bibnamefont {Fakhri}},\ }\bibfield
  {title} {Quantifying dissipation using fluctuating currents,\ }\href
  {https://doi.org/10.1038/s41467-019-09631-x} {\bibfield  {journal} {\bibinfo
  {journal} {Nat. Commun.}\ }\textbf {\bibinfo {volume} {10}},\ \bibinfo
  {pages} {1666} (\bibinfo {year} {2019})}\BibitemShut {NoStop}%
\bibitem [{\citenamefont {Van~Vu}\ \emph {et~al.}(2020)\citenamefont {Van~Vu},
  \citenamefont {Vo},\ and\ \citenamefont {Hasegawa}}]{Vu2020Entropy}%
  \BibitemOpen
  \bibfield  {author} {\bibinfo {author} {\bibfnamefont {T.}~\bibnamefont
  {Van~Vu}}, \bibinfo {author} {\bibfnamefont {V.~T.}\ \bibnamefont {Vo}},\
  and\ \bibinfo {author} {\bibfnamefont {Y.}~\bibnamefont {Hasegawa}},\
  }\bibfield  {title} {Entropy production estimation with optimal current,\
  }\href {https://doi.org/10.1103/PhysRevE.101.042138} {\bibfield  {journal}
  {\bibinfo  {journal} {Phys. Rev. E}\ }\textbf {\bibinfo {volume} {101}},\
  \bibinfo {pages} {042138} (\bibinfo {year} {2020})}\BibitemShut {NoStop}%
\bibitem [{\citenamefont {Otsubo}\ \emph {et~al.}(2020)\citenamefont {Otsubo},
  \citenamefont {Ito}, \citenamefont {Dechant},\ and\ \citenamefont
  {Sagawa}}]{otsubo2020estimating}%
  \BibitemOpen
  \bibfield  {author} {\bibinfo {author} {\bibfnamefont {S.}~\bibnamefont
  {Otsubo}}, \bibinfo {author} {\bibfnamefont {S.}~\bibnamefont {Ito}},
  \bibinfo {author} {\bibfnamefont {A.}~\bibnamefont {Dechant}},\ and\ \bibinfo
  {author} {\bibfnamefont {T.}~\bibnamefont {Sagawa}},\ }\bibfield  {title}
  {Estimating entropy production by machine learning of short-time fluctuating
  currents,\ }\href {https://doi.org/10.1103/PhysRevE.101.062106} {\bibfield
  {journal} {\bibinfo  {journal} {Phys. Rev. E}\ }\textbf {\bibinfo {volume}
  {101}},\ \bibinfo {pages} {062106} (\bibinfo {year} {2020})}\BibitemShut
  {NoStop}%
\bibitem [{\citenamefont {Dechant}\ and\ \citenamefont
  {Sasa}()}]{dechant2021improving}%
  \BibitemOpen
  \bibfield  {author} {\bibinfo {author} {\bibfnamefont {A.}~\bibnamefont
  {Dechant}}\ and\ \bibinfo {author} {\bibfnamefont {S.-i.}\ \bibnamefont
  {Sasa}},\ }\bibfield  {title} {Improving thermodynamic bounds using
  correlations,\ }\href {https://arxiv.org/abs/2104.04169} {\bibinfo  {journal}
  {arXiv:2104.04169}\ }\BibitemShut {NoStop}%
\bibitem [{\citenamefont {Kim}\ \emph {et~al.}(2020)\citenamefont {Kim},
  \citenamefont {Bae}, \citenamefont {Lee},\ and\ \citenamefont
  {Jeong}}]{kim2020neep}%
  \BibitemOpen
\bibfield  {journal} {  }\bibfield  {author} {\bibinfo {author} {\bibfnamefont
  {D.-K.}\ \bibnamefont {Kim}}, \bibinfo {author} {\bibfnamefont
  {Y.}~\bibnamefont {Bae}}, \bibinfo {author} {\bibfnamefont {S.}~\bibnamefont
  {Lee}},\ and\ \bibinfo {author} {\bibfnamefont {H.}~\bibnamefont {Jeong}},\
  }\bibfield  {title} {Learning Entropy Production via Neural Networks,\ }\href
  {https://doi.org/10.1103/PhysRevLett.125.140604} {\bibfield  {journal}
  {\bibinfo  {journal} {Phys. Rev. Lett.}\ }\textbf {\bibinfo {volume} {125}},\
  \bibinfo {pages} {140604} (\bibinfo {year} {2020})}\BibitemShut {NoStop}%
\bibitem [{\citenamefont {Otsubo}\ \emph {et~al.}()\citenamefont {Otsubo},
  \citenamefont {Manikandan}, \citenamefont {Sagawa},\ and\ \citenamefont
  {Krishnamurthy}}]{otsubo2020time}%
  \BibitemOpen
  \bibfield  {author} {\bibinfo {author} {\bibfnamefont {S.}~\bibnamefont
  {Otsubo}}, \bibinfo {author} {\bibfnamefont {S.~K.}\ \bibnamefont
  {Manikandan}}, \bibinfo {author} {\bibfnamefont {T.}~\bibnamefont {Sagawa}},\
  and\ \bibinfo {author} {\bibfnamefont {S.}~\bibnamefont {Krishnamurthy}},\
  }\bibfield  {title} {Estimating time-dependent entropy production from
  non-equilibrium trajectories,\ }\href {https://arxiv.org/abs/2010.03852}
  {\bibinfo  {journal} {arXiv:2010.03852}\ }\BibitemShut {NoStop}%
\bibitem [{\citenamefont {Bae}\ \emph {et~al.}()\citenamefont {Bae},
  \citenamefont {Kim},\ and\ \citenamefont {Jeong}}]{bae2021attaining}%
  \BibitemOpen
\bibfield  {journal} {  }\bibfield  {author} {\bibinfo {author} {\bibfnamefont
  {Y.}~\bibnamefont {Bae}}, \bibinfo {author} {\bibfnamefont {D.-K.}\
  \bibnamefont {Kim}},\ and\ \bibinfo {author} {\bibfnamefont {H.}~\bibnamefont
  {Jeong}},\ }\bibfield  {title} {Attaining entropy production and dissipation
  maps from Brownian movies via neural networks,\ }\href
  {https://arxiv.org/abs/2106.15108} {\bibinfo  {journal} {arXiv:2106.15108}\
  }\BibitemShut {NoStop}%
\bibitem [{\citenamefont {Donsker}\ and\ \citenamefont
  {Varadhan}(2020)}]{donsker1983varadhan}%
  \BibitemOpen
\bibfield  {journal} {  }\bibfield  {author} {\bibinfo {author} {\bibfnamefont
  {M.~D.}\ \bibnamefont {Donsker}}\ and\ \bibinfo {author} {\bibfnamefont
  {S.~R.~S.}\ \bibnamefont {Varadhan}},\ }\bibfield  {title} {Asymptotic
  evaluation of certain markov process expectations for large time. IV,\ }\href
  {https://doi.org/10.1002/cpa.3160360204} {\bibfield  {journal} {\bibinfo
  {journal} {Commun. Pure Appl. Math.}\ }\textbf {\bibinfo {volume} {36}},\
  \bibinfo {pages} {183} (\bibinfo {year} {2020})}\BibitemShut {NoStop}%
\bibitem [{\citenamefont {Spinney}\ and\ \citenamefont
  {Ford}(2012)}]{Richard2012Nonequilibrium}%
  \BibitemOpen
  \bibfield  {author} {\bibinfo {author} {\bibfnamefont {R.~E.}\ \bibnamefont
  {Spinney}}\ and\ \bibinfo {author} {\bibfnamefont {I.~J.}\ \bibnamefont
  {Ford}},\ }\bibfield  {title} {Nonequilibrium Thermodynamics of Stochastic
  Systems with Odd and Even Variables,\ }\href
  {https://doi.org/10.1103/PhysRevLett.108.170603} {\bibfield  {journal}
  {\bibinfo  {journal} {Phys. Rev. Lett.}\ }\textbf {\bibinfo {volume} {108}},\
  \bibinfo {pages} {170603} (\bibinfo {year} {2012})}\BibitemShut {NoStop}%
\bibitem [{\citenamefont {Lee}\ \emph {et~al.}(2013)\citenamefont {Lee},
  \citenamefont {Kwon},\ and\ \citenamefont {Park}}]{lee2013fluctuation}%
  \BibitemOpen
  \bibfield  {author} {\bibinfo {author} {\bibfnamefont {H.~K.}\ \bibnamefont
  {Lee}}, \bibinfo {author} {\bibfnamefont {C.}~\bibnamefont {Kwon}},\ and\
  \bibinfo {author} {\bibfnamefont {H.}~\bibnamefont {Park}},\ }\bibfield
  {title} {Fluctuation Theorems and Entropy Production with Odd-Parity
  Variables,\ }\href {https://doi.org/10.1103/PhysRevLett.110.050602}
  {\bibfield  {journal} {\bibinfo  {journal} {Phys. Rev. Lett.}\ }\textbf
  {\bibinfo {volume} {110}},\ \bibinfo {pages} {050602} (\bibinfo {year}
  {2013})}\BibitemShut {NoStop}%
\bibitem [{\citenamefont {Lee}\ \emph {et~al.}(2021)\citenamefont {Lee},
  \citenamefont {Park},\ and\ \citenamefont {Park}}]{lee2021universal}%
  \BibitemOpen
  \bibfield  {author} {\bibinfo {author} {\bibfnamefont {J.~S.}\ \bibnamefont
  {Lee}}, \bibinfo {author} {\bibfnamefont {J.-M.}\ \bibnamefont {Park}},\ and\
  \bibinfo {author} {\bibfnamefont {H.}~\bibnamefont {Park}},\ }\bibfield
  {title} {Universal form of thermodynamic uncertainty relation for Langevin
  dynamics,\ }\href {https://doi.org/10.1103/PhysRevE.104.L052102} {\bibfield
  {journal} {\bibinfo  {journal} {Phys. Rev. E}\ }\textbf {\bibinfo {volume}
  {104}},\ \bibinfo {pages} {L052102} (\bibinfo {year} {2021})}\BibitemShut
  {NoStop}%
\bibitem [{\citenamefont {Risken}(1991)}]{Risken}%
  \BibitemOpen
  \bibfield  {author} {\bibinfo {author} {\bibfnamefont {H.}~\bibnamefont
  {Risken}},\ }\href@noop {} {\emph {\bibinfo {title} {The Fokker-Planck
  Equation: Methods of Solutions and Applications}}}\ (\bibinfo  {publisher}
  {Springer-Verlag, Berlin},\ \bibinfo {year} {1991})\BibitemShut {NoStop}%
\bibitem [{\citenamefont {Shankar}\ and\ \citenamefont
  {Marchetti}(2018)}]{shankar2018hidden}%
  \BibitemOpen
  \bibfield  {author} {\bibinfo {author} {\bibfnamefont {S.}~\bibnamefont
  {Shankar}}\ and\ \bibinfo {author} {\bibfnamefont {M.~C.}\ \bibnamefont
  {Marchetti}},\ }\bibfield  {title} {Hidden entropy production and work
  fluctuations in an ideal active gas,\ }\href
  {https://doi.org/10.1103/PhysRevE.98.020604} {\bibfield  {journal} {\bibinfo
  {journal} {Phys. Rev. E}\ }\textbf {\bibinfo {volume} {98}},\ \bibinfo
  {pages} {020604} (\bibinfo {year} {2018})}\BibitemShut {NoStop}%
\bibitem [{\citenamefont {Dabelow}\ \emph {et~al.}(2019)\citenamefont
  {Dabelow}, \citenamefont {Bo},\ and\ \citenamefont
  {Eichhorn}}]{Dabelow2019Irreversibility}%
  \BibitemOpen
  \bibfield  {author} {\bibinfo {author} {\bibfnamefont {L.}~\bibnamefont
  {Dabelow}}, \bibinfo {author} {\bibfnamefont {S.}~\bibnamefont {Bo}},\ and\
  \bibinfo {author} {\bibfnamefont {R.}~\bibnamefont {Eichhorn}},\ }\bibfield
  {title} {Irreversibility in Active Matter Systems: Fluctuation Theorem and
  Mutual Information,\ }\href {https://doi.org/10.1103/PhysRevX.9.021009}
  {\bibfield  {journal} {\bibinfo  {journal} {Phys. Rev. X}\ }\textbf {\bibinfo
  {volume} {9}},\ \bibinfo {pages} {021009} (\bibinfo {year}
  {2019})}\BibitemShut {NoStop}%
\bibitem [{\citenamefont {Van~Vu}\ and\ \citenamefont
  {Hasegawa}(2019)}]{vu2019tur}%
  \BibitemOpen
  \bibfield  {author} {\bibinfo {author} {\bibfnamefont {T.}~\bibnamefont
  {Van~Vu}}\ and\ \bibinfo {author} {\bibfnamefont {Y.}~\bibnamefont
  {Hasegawa}},\ }\bibfield  {title} {Uncertainty relations for underdamped
  Langevin dynamics,\ }\href {https://doi.org/10.1103/PhysRevE.100.032130}
  {\bibfield  {journal} {\bibinfo  {journal} {Phys. Rev. E}\ }\textbf {\bibinfo
  {volume} {100}},\ \bibinfo {pages} {032130} (\bibinfo {year}
  {2019})}\BibitemShut {NoStop}%
\bibitem [{\citenamefont {Lee}\ \emph {et~al.}(2019)\citenamefont {Lee},
  \citenamefont {Park},\ and\ \citenamefont {Park}}]{Lee2019Thermodynamic}%
  \BibitemOpen
  \bibfield  {author} {\bibinfo {author} {\bibfnamefont {J.~S.}\ \bibnamefont
  {Lee}}, \bibinfo {author} {\bibfnamefont {J.-M.}\ \bibnamefont {Park}},\ and\
  \bibinfo {author} {\bibfnamefont {H.}~\bibnamefont {Park}},\ }\bibfield
  {title} {Thermodynamic uncertainty relation for underdamped Langevin systems
  driven by a velocity-dependent force,\ }\href
  {https://doi.org/10.1103/PhysRevE.100.062132} {\bibfield  {journal} {\bibinfo
   {journal} {Phys. Rev. E}\ }\textbf {\bibinfo {volume} {100}},\ \bibinfo
  {pages} {062132} (\bibinfo {year} {2019})}\BibitemShut {NoStop}%
\bibitem [{Note1()}]{Note1}%
  \BibitemOpen
  \bibinfo {note} {Here, we note that $[\protect \mathsf {\Gamma }]_{ij}$ is
  not $\delta _{i,j}\gamma /m$ but $\delta _{i,j}\gamma $. This means that the
  governing equation is not reduced to an overdamped Langevin equation when
  mass $m$ approaches zero because the velocity autocorrelation is proportional
  to $e^{-\tau \gamma }$ rather than $e^{-\tau \gamma /m}$.}\BibitemShut
  {Stop}%
\bibitem [{\citenamefont {Vanden-Eijnden}\ and\ \citenamefont
  {Ciccotti}(2006)}]{vanden2006second}%
  \BibitemOpen
  \bibfield  {author} {\bibinfo {author} {\bibfnamefont {E.}~\bibnamefont
  {Vanden-Eijnden}}\ and\ \bibinfo {author} {\bibfnamefont {G.}~\bibnamefont
  {Ciccotti}},\ }\bibfield  {title} {Second-order integrators for Langevin
  equations with holonomic constraints,\ }\href
  {https://doi.org/10.1016/j.cplett.2006.07.086} {\bibfield  {journal}
  {\bibinfo  {journal} {Chemical physics letters}\ }\textbf {\bibinfo {volume}
  {429}},\ \bibinfo {pages} {310} (\bibinfo {year} {2006})}\BibitemShut
  {NoStop}%
\bibitem [{\citenamefont {Seifert}(2012)}]{seifert2012stochastic}%
  \BibitemOpen
  \bibfield  {author} {\bibinfo {author} {\bibfnamefont {U.}~\bibnamefont
  {Seifert}},\ }\bibfield  {title} {Stochastic thermodynamics, fluctuation
  theorems and molecular machines,\ }\href
  {https://doi.org/10.1088/0034-4885/75/12/126001} {\bibfield  {journal}
  {\bibinfo  {journal} {Rep. Prog. Phys.}\ }\textbf {\bibinfo {volume} {75}},\
  \bibinfo {pages} {126001} (\bibinfo {year} {2012})}\BibitemShut {NoStop}%
\bibitem [{\citenamefont {Kingma}\ and\ \citenamefont {Ba}()}]{kingma2014adam}%
  \BibitemOpen
  \bibfield  {author} {\bibinfo {author} {\bibfnamefont {D.~P.}\ \bibnamefont
  {Kingma}}\ and\ \bibinfo {author} {\bibfnamefont {J.}~\bibnamefont {Ba}},\
  }\bibfield  {title} {Adam: A Method for Stochastic Optimization,\ }\href
  {https://arxiv.org/abs/1412.6980} {\bibinfo  {journal} {arXiv:1412.6980}\
  }\BibitemShut {NoStop}%
\bibitem [{\citenamefont {Saxe}\ \emph {et~al.}(2014)\citenamefont {Saxe},
  \citenamefont {McClelland},\ and\ \citenamefont {Ganguli}}]{saxe2014ortho}%
  \BibitemOpen
\bibfield  {journal} {  }\bibfield  {author} {\bibinfo {author} {\bibfnamefont
  {A.~M.}\ \bibnamefont {Saxe}}, \bibinfo {author} {\bibfnamefont {J.~L.}\
  \bibnamefont {McClelland}},\ and\ \bibinfo {author} {\bibfnamefont
  {S.}~\bibnamefont {Ganguli}},\ }\bibfield  {title} {Exact solutions to the
  nonlinear dynamics of learning in deep linear neural networks,\ }in\ \href
  {https://openreview.net/forum?id=_wzZwKpTDF_9C} {\emph {\bibinfo {booktitle}
  {International Conference on Learning Representations}}}\ (\bibinfo {year}
  {2014})\BibitemShut {NoStop}%
\bibitem [{\citenamefont {Mart{\'\i}nez}\ \emph {et~al.}(2019)\citenamefont
  {Mart{\'\i}nez}, \citenamefont {Bisker}, \citenamefont {Horowitz},\ and\
  \citenamefont {Parrondo}}]{martinez2019inferring}%
  \BibitemOpen
  \bibfield  {author} {\bibinfo {author} {\bibfnamefont {I.~A.}\ \bibnamefont
  {Mart{\'\i}nez}}, \bibinfo {author} {\bibfnamefont {G.}~\bibnamefont
  {Bisker}}, \bibinfo {author} {\bibfnamefont {J.~M.}\ \bibnamefont
  {Horowitz}},\ and\ \bibinfo {author} {\bibfnamefont {J.~M.}\ \bibnamefont
  {Parrondo}},\ }\bibfield  {title} {Inferring broken detailed balance in the
  absence of observable currents,\ }\href
  {https://doi.org/10.1038/s41467-019-11051-w} {\bibfield  {journal} {\bibinfo
  {journal} {Nat. Commun.}\ }\textbf {\bibinfo {volume} {10}},\ \bibinfo
  {pages} {3542} (\bibinfo {year} {2019})}\BibitemShut {NoStop}%
\bibitem [{\citenamefont {Gillespie}(1977)}]{gillespie1977exact}%
  \BibitemOpen
  \bibfield  {author} {\bibinfo {author} {\bibfnamefont {D.~T.}\ \bibnamefont
  {Gillespie}},\ }\bibfield  {title} {Exact stochastic simulation of coupled
  chemical reactions,\ }\href {https://doi.org/10.1021/j100540a008} {\bibfield
  {journal} {\bibinfo  {journal} {J. Phys. Chem.}\ }\textbf {\bibinfo {volume}
  {81}},\ \bibinfo {pages} {2340} (\bibinfo {year} {1977})}\BibitemShut
  {NoStop}%
\bibitem [{\citenamefont {Skinner}\ and\ \citenamefont
  {Dunkel}(2021)}]{skinner2021wtd}%
  \BibitemOpen
  \bibfield  {author} {\bibinfo {author} {\bibfnamefont {D.~J.}\ \bibnamefont
  {Skinner}}\ and\ \bibinfo {author} {\bibfnamefont {J.}~\bibnamefont
  {Dunkel}},\ }\bibfield  {title} {Estimating Entropy Production from Waiting
  Time Distributions,\ }\href {https://doi.org/10.1103/PhysRevLett.127.198101}
  {\bibfield  {journal} {\bibinfo  {journal} {Phys. Rev. Lett.}\ }\textbf
  {\bibinfo {volume} {127}},\ \bibinfo {pages} {198101} (\bibinfo {year}
  {2021})}\BibitemShut {NoStop}%
\bibitem [{\citenamefont {Kawaguchi}\ and\ \citenamefont
  {Nakayama}(2013)}]{kawaguchi2013hidden}%
  \BibitemOpen
  \bibfield  {author} {\bibinfo {author} {\bibfnamefont {K.}~\bibnamefont
  {Kawaguchi}}\ and\ \bibinfo {author} {\bibfnamefont {Y.}~\bibnamefont
  {Nakayama}},\ }\bibfield  {title} {Fluctuation theorem for hidden entropy
  production,\ }\href {https://doi.org/10.1103/PhysRevE.88.022147} {\bibfield
  {journal} {\bibinfo  {journal} {Phys. Rev. E}\ }\textbf {\bibinfo {volume}
  {88}},\ \bibinfo {pages} {022147} (\bibinfo {year} {2013})}\BibitemShut
  {NoStop}%
\bibitem [{\citenamefont {Esposito}\ and\ \citenamefont
  {Parrondo}(2015)}]{esposito2015hidden}%
  \BibitemOpen
  \bibfield  {author} {\bibinfo {author} {\bibfnamefont {M.}~\bibnamefont
  {Esposito}}\ and\ \bibinfo {author} {\bibfnamefont {J.~M.~R.}\ \bibnamefont
  {Parrondo}},\ }\bibfield  {title} {Stochastic thermodynamics of hidden
  pumps,\ }\href {https://doi.org/10.1103/PhysRevE.91.052114} {\bibfield
  {journal} {\bibinfo  {journal} {Phys. Rev. E}\ }\textbf {\bibinfo {volume}
  {91}},\ \bibinfo {pages} {052114} (\bibinfo {year} {2015})}\BibitemShut
  {NoStop}%
\bibitem [{\citenamefont {Paszke}\ \emph {et~al.}(2019)\citenamefont {Paszke},
  \citenamefont {Gross}, \citenamefont {Massa}, \citenamefont {Lerer},
  \citenamefont {Bradbury}, \citenamefont {Chanan}, \citenamefont {Killeen},
  \citenamefont {Lin}, \citenamefont {Gimelshein}, \citenamefont {Antiga},
  \citenamefont {Desmaison}, \citenamefont {Kopf}, \citenamefont {Yang},
  \citenamefont {DeVito}, \citenamefont {Raison}, \citenamefont {Tejani},
  \citenamefont {Chilamkurthy}, \citenamefont {Steiner}, \citenamefont {Fang},
  \citenamefont {Bai},\ and\ \citenamefont {Chintala}}]{pytorch2019}%
  \BibitemOpen
  \bibfield  {author} {\bibinfo {author} {\bibfnamefont {A.}~\bibnamefont
  {Paszke}}, \bibinfo {author} {\bibfnamefont {S.}~\bibnamefont {Gross}},
  \bibinfo {author} {\bibfnamefont {F.}~\bibnamefont {Massa}}, \bibinfo
  {author} {\bibfnamefont {A.}~\bibnamefont {Lerer}}, \bibinfo {author}
  {\bibfnamefont {J.}~\bibnamefont {Bradbury}}, \bibinfo {author}
  {\bibfnamefont {G.}~\bibnamefont {Chanan}}, \bibinfo {author} {\bibfnamefont
  {T.}~\bibnamefont {Killeen}}, \bibinfo {author} {\bibfnamefont
  {Z.}~\bibnamefont {Lin}}, \bibinfo {author} {\bibfnamefont {N.}~\bibnamefont
  {Gimelshein}}, \bibinfo {author} {\bibfnamefont {L.}~\bibnamefont {Antiga}},
  \bibinfo {author} {\bibfnamefont {A.}~\bibnamefont {Desmaison}}, \bibinfo
  {author} {\bibfnamefont {A.}~\bibnamefont {Kopf}}, \bibinfo {author}
  {\bibfnamefont {E.}~\bibnamefont {Yang}}, \bibinfo {author} {\bibfnamefont
  {Z.}~\bibnamefont {DeVito}}, \bibinfo {author} {\bibfnamefont
  {M.}~\bibnamefont {Raison}}, \bibinfo {author} {\bibfnamefont
  {A.}~\bibnamefont {Tejani}}, \bibinfo {author} {\bibfnamefont
  {S.}~\bibnamefont {Chilamkurthy}}, \bibinfo {author} {\bibfnamefont
  {B.}~\bibnamefont {Steiner}}, \bibinfo {author} {\bibfnamefont
  {L.}~\bibnamefont {Fang}}, \bibinfo {author} {\bibfnamefont {J.}~\bibnamefont
  {Bai}},\ and\ \bibinfo {author} {\bibfnamefont {S.}~\bibnamefont
  {Chintala}},\ }\bibfield  {title} {PyTorch: An Imperative Style,
  High-Performance Deep Learning Library,\ }in\ \href
  {http://papers.neurips.cc/paper/9015-pytorch-an-imperative-style-high-performance-deep-learning-library.pdf}
  {\emph {\bibinfo {booktitle} {Advances in Neural Information Processing
  Systems 32}}}\ (\bibinfo  {publisher} {Curran Associates, Inc.},\ \bibinfo
  {year} {2019})\ pp.\ \bibinfo {pages} {8024--8035}\BibitemShut {NoStop}%
\bibitem [{Note2()}]{Note2}%
  \BibitemOpen
  \bibinfo {note} {\protect \url
  {https://github.com/kdkyum/odd_neep}}\BibitemShut {NoStop}%
\end{thebibliography}
\end{document}